
\input amssym.def


\magnification=\magstephalf
\hsize=14.0 true cm
\vsize=19 true cm
\hoffset=1.0 true cm
\voffset=2.0 true cm

\abovedisplayskip=12pt plus 3pt minus 3pt
\belowdisplayskip=12pt plus 3pt minus 3pt
\parindent=1em


\font\sixrm=cmr6
\font\eightrm=cmr8
\font\ninerm=cmr9

\font\sixi=cmmi6
\font\eighti=cmmi8
\font\ninei=cmmi9

\font\sixsy=cmsy6
\font\eightsy=cmsy8
\font\ninesy=cmsy9

\font\sixbf=cmbx6
\font\eightbf=cmbx8
\font\ninebf=cmbx9

\font\eightit=cmti8
\font\nineit=cmti9

\font\eightsl=cmsl8
\font\ninesl=cmsl9

\font\sixss=cmss8 at 8 true pt
\font\sevenss=cmss9 at 9 true pt
\font\eightss=cmss8
\font\niness=cmss9
\font\tenss=cmss10

 at 12 true pt
\font\bigrm=cmr10 at 12 true pt
\font\bigbf=cmbx10 at 12 true pt

 at 16 true pt
 at 16 true pt
 at 16 true pt

\catcode`@=11
\newfam\ssfam

\def\tenpoint{\def\rm{\fam0\tenrm}%
    \textfont0=\tenrm \scriptfont0=\sevenrm \scriptscriptfont0=\fiverm
    \textfont1=\teni  \scriptfont1=\seveni  \scriptscriptfont1=\fivei
    \textfont2=\tensy \scriptfont2=\sevensy \scriptscriptfont2=\fivesy
    \textfont3=\tenex \scriptfont3=\tenex   \scriptscriptfont3=\tenex
    \textfont\itfam=\tenit                  \def\it{\fam\itfam\tenit}%
    \textfont\slfam=\tensl                  \def\sl{\fam\slfam\tensl}%
    \textfont\bffam=\tenbf \scriptfont\bffam=\sevenbf
    \scriptscriptfont\bffam=\fivebf
                                            \def\bf{\fam\bffam\tenbf}%
    \textfont\ssfam=\tenss \scriptfont\ssfam=\sevenss
    \scriptscriptfont\ssfam=\sevenss
                                            \def\ss{\fam\ssfam\tenss}%
    \normalbaselineskip=13pt
    \setbox\strutbox=\hbox{\vrule height8.5pt depth3.5pt width0pt}%
    \let\big=\tenbig
    \normalbaselines\rm}

\def\ninepoint{\def\rm{\fam0\ninerm}%
    \textfont0=\ninerm      \scriptfont0=\sixrm
                            \scriptscriptfont0=\fiverm
    \textfont1=\ninei       \scriptfont1=\sixi
                            \scriptscriptfont1=\fivei
    \textfont2=\ninesy      \scriptfont2=\sixsy
                            \scriptscriptfont2=\fivesy
    \textfont3=\tenex       \scriptfont3=\tenex
                            \scriptscriptfont3=\tenex
    \textfont\itfam=\nineit \def\it{\fam\itfam\nineit}%
    \textfont\slfam=\ninesl \def\sl{\fam\slfam\ninesl}%
    \textfont\bffam=\ninebf \scriptfont\bffam=\sixbf
                            \scriptscriptfont\bffam=\fivebf
                            \def\bf{\fam\bffam\ninebf}%
    \textfont\ssfam=\niness \scriptfont\ssfam=\sixss
                            \scriptscriptfont\ssfam=\sixss
                            \def\ss{\fam\ssfam\niness}%
    \normalbaselineskip=12pt
    \setbox\strutbox=\hbox{\vrule height8.0pt depth3.0pt width0pt}%
    \let\big=\ninebig
    \normalbaselines\rm}

\def\eightpoint{\def\rm{\fam0\eightrm}%
    \textfont0=\eightrm      \scriptfont0=\sixrm
                             \scriptscriptfont0=\fiverm
    \textfont1=\eighti       \scriptfont1=\sixi
                             \scriptscriptfont1=\fivei
    \textfont2=\eightsy      \scriptfont2=\sixsy
                             \scriptscriptfont2=\fivesy
    \textfont3=\tenex        \scriptfont3=\tenex
                             \scriptscriptfont3=\tenex
    \textfont\itfam=\eightit \def\it{\fam\itfam\eightit}%
    \textfont\slfam=\eightsl \def\sl{\fam\slfam\eightsl}%
    \textfont\bffam=\eightbf \scriptfont\bffam=\sixbf
                             \scriptscriptfont\bffam=\fivebf
                             \def\bf{\fam\bffam\eightbf}%
    \textfont\ssfam=\eightss \scriptfont\ssfam=\sixss
                             \scriptscriptfont\ssfam=\sixss
                             \def\ss{\fam\ssfam\eightss}%
    \normalbaselineskip=10pt
    \setbox\strutbox=\hbox{\vrule height7.0pt depth2.0pt width0pt}%
    \let\big=\eightbig
    \normalbaselines\rm}

\def\tenbig#1{{\hbox{$\left#1\vbox to8.5pt{}\right.\n@space$}}}
\def\ninebig#1{{\hbox{$\textfont0=\tenrm\textfont2=\tensy
                       \left#1\vbox to7.25pt{}\right.\n@space$}}}
\def\eightbig#1{{\hbox{$\textfont0=\ninerm\textfont2=\ninesy
                       \left#1\vbox to6.5pt{}\right.\n@space$}}}

\font\sectionfont=cmbx10
\font\subsectionfont=cmti10

\def\figurecaptionfont{\ninepoint}
\def\tablecaptionfont{\ninepoint}
\def\footnotefont{\eightpoint}


\newcount\equationno
\newcount\bibitemno
\newcount\figureno
\newcount\tableno

\equationno=0
\bibitemno=0
\figureno=0
\tableno=0
\advance\pageno by -1


\footline={\ifnum\pageno=0{\hfil}\else
{\hss\rm\the\pageno\hss}\fi}


\def\section #1. #2 \par
{\vskip0pt plus .20\vsize\penalty-100 \vskip0pt plus-.20\vsize
\vskip 1.6 true cm plus 0.2 true cm minus 0.2 true cm
\global\def\equationlabel{#1}
\global\equationno=0
\leftline{\sectionfont #1. #2}\par
\immediate\write\terminal{Section #1. #2}
\vskip 0.7 true cm plus 0.1 true cm minus 0.1 true cm
\noindent}


\def\subsection #1 \par
{\vskip0pt plus 0.8 true cm\penalty-50 \vskip0pt plus-0.8 true cm
\vskip2.5ex plus 0.1ex minus 0.1ex
\leftline{\subsectionfont #1}\par
\immediate\write\terminal{Subsection #1}
\vskip1.0ex plus 0.1ex minus 0.1ex
\noindent}


\def\appendix #1 \par
{\vskip0pt plus .20\vsize\penalty-100 \vskip0pt plus-.20\vsize
\vskip 1.6 true cm plus 0.2 true cm minus 0.2 true cm
\global\def\equationlabel{\hbox{\rm#1}}
\global\equationno=0
\leftline{\sectionfont Appendix #1}\par
\immediate\write\terminal{Appendix #1}
\vskip 0.7 true cm plus 0.1 true cm minus 0.1 true cm
\noindent}


\def\enum{\global\advance\equationno by 1
(\equationlabel.\the\equationno)}


\def\ifundefined#1{\expandafter\ifx\csname#1\endcsname\relax}

\def\ref#1{\ifundefined{#1}?\immediate\write\terminal{unknown reference
on page \the\pageno}\else\csname#1\endcsname\fi}

\newwrite\terminal
\newwrite\bibitemlist

\def\bibitem#1#2\par{\global\advance\bibitemno by 1
\immediate\write\bibitemlist{\string\def
\expandafter\string\csname#1\endcsname
{\the\bibitemno}}
\item{[\the\bibitemno]}#2\par}

\def\beginbibliography{
\vskip0pt plus .15\vsize\penalty-100 \vskip0pt plus-.15\vsize
\vskip 1.2 true cm plus 0.2 true cm minus 0.2 true cm
\leftline{\sectionfont References}\par
\immediate\write\terminal{References}
\immediate\openout\bibitemlist=biblist
\frenchspacing\parindent=1.5em
\vskip 0.5 true cm plus 0.1 true cm minus 0.1 true cm}

\def\endbibliography{
\immediate\closeout\bibitemlist
\nonfrenchspacing\parindent=1.0em}


\def\figurecaption#1{\global\advance\figureno by 1
\narrower\figurecaptionfont
Fig.~\the\figureno. #1}

\def\tablecaption#1{\global\advance\tableno by 1
\vbox to 0.5 true cm { }
\centerline{\tablecaptionfont%
Table~\the\tableno. #1}
\vskip-0.4 true cm}

\tenpoint

\immediate\openin\bibitemlist=biblist
\ifeof\bibitemlist\immediate\closein\bibitemlist
\else\immediate\closein\bibitemlist
\input biblist \fi


\def\thisyear{\number\year}

\def\thismonth{\ifcase\month\or
January\or February\or March\or April\or May\or June\or
July\or August\or September\or October\or November\or December\fi}



\def\rmd{{\rm d}}

\def\rme{{\rm e}}


\def\rz{{\Bbb R}}
\def\gz{{\Bbb Z}}


\def\proof{\noindent{\sl Proof:}\kern0.6em}
\def\endproof{\hskip0.6em plus0.1em minus0.1em
\setbox0=\null\ht0=5.4pt\dp0=1pt\wd0=5.3pt
\vbox{\hrule height0.8pt
\hbox{\vrule width0.8pt\box0\vrule width0.8pt}
\hrule height0.8pt}}
\def\frac#1#2{\hbox{$#1\over#2$}}
\def\dual{\mathstrut^*\kern-0.1em}

\def\lvec#1{\setbox0=\hbox{$#1$}
    \setbox1=\hbox{$\scriptstyle\leftarrow$}
    #1\kern-\wd0\smash{
    \raise\ht0\hbox{$\raise1pt\hbox{$\scriptstyle\leftarrow$}$}}
    \kern-\wd1\kern\wd0}
\def\rvec#1{\setbox0=\hbox{$#1$}
    \setbox1=\hbox{$\scriptstyle\rightarrow$}
    #1\kern-\wd0\smash{
    \raise\ht0\hbox{$\raise1pt\hbox{$\scriptstyle\rightarrow$}$}}
    \kern-\wd1\kern\wd0}


\def\nabstar#1{\nabla\kern-0.5pt\smash{\raise 4.5pt\hbox{$\ast$}}
               \kern-4.5pt_{#1}}
\def\drv#1{{\partial_{#1}}}
\def\drvstar#1{\partial\kern-0.5pt\smash{\raise 4.5pt\hbox{$\ast$}}
               \kern-5.0pt_{#1}}
\def\ldrv#1{{\lvec{\,\partial}_{#1}}}
\def\ldrvstar#1{\lvec{\,\partial}\kern-0.5pt\smash{\raise 4.5pt\hbox{$\ast$}}
               \kern-5.0pt_{#1}}




\def\block{{\Bbb B}}


\def\dirac#1{\gamma_{#1}}
\def\diracstar#1#2{
    \setbox0=\hbox{$\gamma$}\setbox1=\hbox{$\gamma_{#1}$}
    \gamma_{#1}\kern-\wd1\kern\wd0
    \smash{\raise4.5pt\hbox{$\scriptstyle#2$}}}


\def\SUtwo{{\rm SU(2)}}

\def\tr{{\rm tr}\,}


\def\A{\alpha}
\def\B{\beta}
\def\C{\gamma}


\def\dx#1{\rmd x_{#1}}
\def\Om#1{\Omega_{#1}}
\rightline{DESY 98-095}

\vskip 3.0 true cm minus 0.3 true cm
\centerline
{\bigbf Topology and the axial anomaly in abelian}
\vskip 1.5ex
\centerline
{\bigbf lattice gauge theories}
\vskip 1.5 true cm
\centerline{\bigrm Martin L\"uscher}
\vskip1ex
\centerline{\it Deutsches Elektronen-Synchrotron DESY}
\centerline{\it Notkestrasse 85, D-22603 Hamburg, Germany}
\centerline{\it E-mail: luscher@mail.desy.de}
\vskip 2.5 true cm
\centerline{\bf Abstract}
\vskip 2.0ex
The axial anomaly in abelian lattice gauge theories is shown to
be equal to a simple quadratic expression in the gauge field tensor
plus a removable divergence term
if the lattice Dirac operator satisfies the Ginsparg-Wilson relation.
The theorem is a consequence of the locality, the 
gauge invariance and the topological nature of the anomaly
and does not refer to any other properties of the lattice theory. 
\vfill
\centerline{\thismonth\space\thisyear}
\eject

\section 1. Introduction

Chiral gauge theories with anomaly-free multiplets of Weyl fermions
appear to be well-defined quantum field theories in the 
continuum limit, but so far have not found a completely
satisfactory formulation on the lattice. 
The recent discovery that chiral symmetry can be exactly preserved on 
the lattice, without doubling of the number of fermion species
or other undesirable features, now 
gives rise to renewed hopes that one might be able to 
solve this long-standing theoretical problem
[\ref{HasenfratzI}--\ref{KikukawaYamada}].

When a ${\rm U}(1)$ gauge field is coupled
to a multiplet of $N$ left-handed Weyl fermions with charges $\rme_{\alpha}$,
the condition for anomaly cancellation is
$$
  \sum_{\alpha=1}^N\rme_{\alpha}^3=0.
  \eqno\enum
$$
In the continuum limit the cancellation of the gauge anomaly
is then a consequence of the fact that the anomaly of the axial
current of a single Dirac fermion is proportional to the square
of its charge.
The situation on the lattice is however not quite as simple
and it is unclear whether 
the anomalies cancel before taking the continuum limit.

In this paper we consider a Dirac fermion coupled to 
an external ${\rm U}(1)$ lattice gauge field and determine the 
general form of the anomaly of the associated axial current.
To ensure the chiral invariance of the fermion action
we assume that the lattice Dirac operator satisfies
the Ginsparg-Wilson relation [\ref{GinspargWilson}].
The axial anomaly then arises 
from the non-invariance of the fermion integration measure 
under chiral transformations [\ref{LuscherI}]. Explicitly one finds that 
the anomaly is proportional to
[\ref{HasenfratzEtAl},\ref{LuscherI}]
$$
  q(x)=-\frac{1}{2}\,\tr\{\dirac{5}D(x,x)\},
  \eqno\enum
$$
where $D(x,y)$ denotes 
the kernel representing the Dirac operator in position space%
~\footnote{$\dagger$}{\footnotefont
The notational conventions used in this paper are summarized in appendix A}.
From eq.~(1.2) and the properties of the Dirac operator one infers
that $q(x)$ is a gauge invariant local field.
Moreover its variation under local deformations of the gauge field satisfies
$$
  \sum_x\,\delta q(x)=0,
  \eqno\enum
$$
which reflects the topological nature of the anomaly.
The exact index theorem of ref.~[\ref{HasenfratzEtAl}]
is another manifestation of this.

The main result obtained here is that any
field $q(x)$ with the properties listed above is of the form
$$
  q(x)=\A+\B_{\mu\nu}F_{\mu\nu}(x)+
  \C\epsilon_{\mu\nu\rho\sigma}
  F_{\mu\nu}(x)F_{\rho\sigma}(x+\hat{\mu}+\hat{\nu})
  +\drvstar{\mu}k_{\mu}(x),
  \eqno\enum
$$
where $F_{\mu\nu}(x)$ denotes the gauge field tensor and
$k_{\mu}(x)$ a gauge invariant local current. 
The precise conditions under which this theorem holds 
will be specified later.
In the case of the anomaly,
the first two terms on the right-hand side of eq.~(1.4)
vanish as a consequence of the lattice symmetries
and the divergence term can be removed through a redefinition of 
the axial current. One is then left with the quadratic expression
in the gauge field tensor. In particular, since the field tensor 
scales with the charge of the fermion, the anomaly cancellation 
may be expected work out as in the continuum theory.

The proof of eq.~(1.4) is complicated and 
requires some preparation. In the next section we
introduce the notion of a differential form on the lattice
and define the corresponding exterior difference operator.
One may then show that closed forms are exact (in most cases)
which is referred to as the Poincar\'e lemma.
Some further technical results are established in sects.~3--5,
where we set up the notations for U(1) lattice gauge fields
and define the class of local composite fields 
which is being considered. After that
eq.~(1.4) will have acquired a 
precise mathematical meaning and its proof can be given on 
a few pages (sect.~6).

\section 2. Differential forms and the Poincar\'e lemma on the lattice

The non-trivial topology of manifolds and vector bundles 
is often related to the existence of certain differential
forms whose exterior derivative vanishes but which cannot
be represented as the derivative of some other form.
In mathematics this goes under 
the heading of cohomology and algebraic topology,
an elegant subject whose significance
for the understanding of anomalies in quantum field theory
has long been appreciated (see refs.~[\ref{BottTu},\ref{Bertlmann}] 
for an introduction and references to the original literature).
The basic cohomology problem on $\rz^n$ is solved 
by the Poincar\'e lemma and the aim in this section is to 
formulate and prove this lemma on the lattice.

\subsection 2.1 Differential forms

Let $\gz^n$ be the lattice of integer vectors 
$x=(x_1,\ldots,x_n)$
in $n$ dimensions where $n\geq1$ is left unspecified. 
In the following we will be concerned with tensor fields
$f_{\mu_1\ldots\mu_k}(x)$ on $\gz^n$
that are totally anti-symmetric
in the indices $\mu_1,\ldots,\mu_k$. As in the continuum
the anti-symmetry is conveniently taken into account by introducing
a Grassmann algebra with basis elements
$$
  \dx{1},\ldots,\dx{n},
  \qquad
  \dx{\mu}\dx{\nu}=-\dx{\nu}\dx{\mu}.
  \eqno\enum
$$
If we adopt the Einstein summation convention for tensor indices,
the general $k$-form on $\gz^n$ is then given by
$$
  f(x)={1\over k!}\,f_{\mu_1\ldots\mu_k}(x)\,\dx{\mu_1}\ldots\dx{\mu_k}.
  \eqno\enum
$$
For simplicity attention will be restricted to 
$k$-forms with compact support and the linear space of all these
forms is denoted by $\Om{k}$. 
The Poincar\'e lemma
is however valid for more general classes of forms too 
(exponentially decaying forms for example).

An exterior difference operator 
$\rmd\!:\Om{k}\to\Om{k+1}$ may now be defined through
$$
  \rmd f(x)={1\over k!}\,\drv{\mu}f_{\mu_1\ldots\mu_k}(x)
  \,\dx{\mu}\dx{\mu_1}\ldots\dx{\mu_k},
  \eqno\enum
$$
where $\drv{\mu}$ denotes the forward 
nearest-neighbour difference operator. 
Explicitly for $k=0,1$ and $2$ we have 
$$
  \eqalignno{
  (\rmd f)_{\mu}&=\drv{\mu}f, 
  &\enum\cr\noalign{\vskip1ex}
  (\rmd f)_{\mu\nu}&=\drv{\mu}f_{\nu}-\drv{\nu}f_{\mu},
  &\enum\cr\noalign{\vskip1ex}
  (\rmd f)_{\mu\nu\rho}&=\drv{\mu}f_{\nu\rho}+
                  \drv{\nu}f_{\rho\mu}+
                  \drv{\rho}f_{\mu\nu},
  &\enum\cr}
$$
and similar formulae are obtained for the larger values of $k$.
The sequence ends at $k=n$ where the definition~(2.3) implies
$\rmd f=0$ since there are no non-trivial $n+1$ forms on $\gz^n$.

The associated divergence operator $\rmd^{\ast}:\Om{k}\to\Om{k-1}$
is defined in the obvious way by setting
$\rmd^{\ast}f=0$ if $f$ is a $0$-form and
$$
  \rmd^{\ast}f(x)={1\over(k-1)!}\,
  \drvstar{\mu}f_{\mu\mu_2\ldots\mu_{k}}(x)\dx{\mu_2}\ldots\dx{\mu_k}
  \eqno\enum
$$
in all other cases, where
$\drvstar{\mu}$ is the backward nearest-neighbour difference operator.
With respect to the natural scalar product for tensor fields on $\gz^n$, 
$\rmd^{\ast}$ is equal to minus the adjoint of $\rmd$.

\subsection 2.2 Poincar\'e lemma

It is straightforward to show that $\rmd^2=0$ and the 
difference equation $\rmd f=0$ is hence solved by all forms $f=\rmd g$.
The Poincar\'e lemma
asserts that these are in fact all solutions,
an exception being the $n$-forms where one has a one-dimensional
space of further solutions.
The precise statement is

\proclaim Lemma 2.1.
Let $f$ be a $k$-form which satisfies 
$$
  \hbox{$\rmd f=0$ and\/ 
  $\sum_xf(x)=0$ if\/ $k=n$.}
  \eqno\enum
$$
Then there exists a form $g\in\Om{k-1}$ such that $f=\rmd g$.

\proof
We first show that the lemma holds for $n=1$ and then proceed 
to higher dimensions by induction. 
On a one-dimensional lattice the non-trivial forms are the $0$- and the
$1$-forms. In the first case the equation $\rmd f=0$ implies
$\drv{\mu}f(x)=0$ and since $f(x)$ is required to have compact support,
we conclude that $f=0$ which is what the lemma claims.
In the other case we have $k=n$ and the only condition on $f$
is then that $\sum_xf(x)=0$. It follows from this that 
the $0$-form
$$
  g(x)=\sum_{y_1=-\infty}^{x_1-1}f_1(y)
  \eqno\enum
$$
has compact support and since $\rmd g=f$ we have thus 
proved the lemma for $n=1$.

Let us now assume that $n$ is greater than $1$ and that the lemma
holds in dimension $n-1$. We then decompose the form $f$ according to
$$
  f=u\,\dx{n}+v,
  \eqno\enum
$$
where $u$ and $v$ are elements of $\Om{k-1}$ and $\Om{k}$ respectively
that are independent of $\dx{n}$. If we ignore the dependence
on $x_n$, these forms may be regarded as forms in $n-1$ dimensions.
To avoid any confusion the corresponding exterior difference operator
will be denoted by $\bar{\rmd}$. It is then straightforward to 
show that
$$
  \rmd f=\left\{\bar{\rmd}u+(-1)^k\drv{n}v\right\}\dx{n}+\bar{\rmd}v 
  \eqno\enum
$$
and the equation $\rmd f=0$ hence implies
$$
  \bar{\rmd}u+(-1)^k\drv{n}v=0.
  \eqno\enum
$$
Note that $v=0$ if $k=n$ and the condition (2.8) reduces to 
$\sum_xu(x)=0$.

We now define a form $\bar{u}$ on $\gz^{n-1}$ through
$$
  \bar{u}(x)=\sum_{y_n=-\infty}^{\infty}u(y),
  \qquad
  y=(x_1,\ldots,x_{n-1},y_n).
  \eqno\enum
$$
Evidently $\bar{u}$ is an element of $\Om{k-1}$ and from the above
one infers that it satisfies the premises of the lemma. 
The induction hypothesis thus allows us to conclude that 
$\bar{u}=\bar{\rmd}\bar{g}$ for some form $\bar{g}\in\Om{k-2}$
in $n-1$ dimensions.

Next we introduce a new form $h$ on $\gz^n$ through
$$
  h(x)=(-1)^{k-1}\sum_{y_n=-\infty}^{x_n-1}
  \left\{u(y)-\delta_{y_n,z_n}\bar{u}(x)\right\},
  \eqno\enum
$$
where $y$ is as in eq.~(2.13) and $z$ an arbitrary reference point.
$h$ has compact support and using eq.~(2.12) it 
is straightforward to prove that 
$$
  f(x)=\delta_{x_n,z_n}\bar{u}(x)\dx{n}+\rmd h(x).
  \eqno\enum
$$
We thus conclude that the sum 
$$
  g(x)=\delta_{x_n,z_n}\bar{g}(x)\dx{n}+h(x)
  \eqno\enum
$$ 
is an element of $\Om{k-1}$ such that $f=\rmd g$.
\endproof

\vskip2ex
The construction of the form $g$ presented above is explicit
and follows a simple logical scheme.
The coefficients of $g$ are just some particular linear 
combinations of the coefficients of $f$.
Moreover, as can be easily shown, $g$~is supported
on the same rectangular block of lattice points as~$f$ if we choose
the reference point~$z$ in eqs.~(2.14)--(2.16) to lie in this
region. 
For the proof of eq.~(1.4)
this locality property will be very important.

An equivalent formulation of the Poincar\'e lemma can be
given in terms of the divergence operator $\rmd^{\ast}$.
We omit the proof (which is similar to the one given above)
and simply quote

\proclaim Lemma 2.2.
Let $f$ be a $k$-form which satisfies 
$$
  \hbox{$\rmd^{\ast}f=0$ and\/ 
  $\sum_xf(x)=0$ if\/ $k=0$.}
  \eqno\enum
$$
Then there exists a form $g\in\Om{k+1}$ such that 
$f=\rmd^{\ast}g$.

\section 3. {\rm U(1)} gauge fields on the lattice

In the so-called compact formulation of abelian lattice gauge theories 
the gauge fields are represented by link variables
$$
  U(x,\mu)\in {\rm U(1)},
  \qquad x=(x_1,\ldots,x_4)\in\gz^4, 
  \qquad \mu=1,\ldots,4.
  \eqno\enum
$$
Note that throughout this paper all fields live on the infinite lattice
and no boundary conditions will be imposed.
Gauge transformations $\Lambda(x)$ also take values in 
U(1) and act on the gauge fields through
$$
  U(x,\mu)\to\Lambda(x)U(x,\mu)\Lambda(x+\hat{\mu})^{-1},
  \eqno\enum
$$
where $\hat{\mu}$ denotes the unit vector in direction $\mu$.
Not all gauge fields will be admitted later, but only
those which are smooth at the scale of the lattice spacing
to some degree. 
To make this explicit we introduce the plaquette field
$$
  P(x,\mu,\nu)=U(x,\mu)U(x+\hat{\mu},\nu)
  U(x+\hat{\nu},\mu)^{-1}U(x,\nu)^{-1}
  \eqno\enum
$$
and define the field tensor $F_{\mu\nu}(x)$ through 
$$
  F_{\mu\nu}(x)=
  {1\over i}\ln P(x,\mu,\nu)
  \qquad
  -\pi<F_{\mu\nu}(x)\leq\pi.
  \eqno\enum
$$
Attention is then restricted to the set of fields satisfying
\footnote{$\dagger$}{\footnotefont
When physical units are employed, the right-hand side of 
eq.~(3.5) should be replaced by $\epsilon/a^2$
where $a$ denotes the lattice spacing. It is then immediately
clear that the bound becomes irrelevant in the classical continuum limit}
$$
  \sup_{x,\mu,\nu}\,|F_{\mu\nu}(x)|<\epsilon
  \eqno\enum
$$
for some fixed number $\epsilon$ 
in the range $0<\epsilon<\frac{1}{3}\pi$. 
In the following such fields are referred to as {\it admissible.}

Evidently it would be more appealing if one could do without
such a constraint, but a moment of thought reveals that 
eq.~(1.4) cannot be true for all fields, because
there exists a topological invariant, the magnetic monopole current 
$\epsilon_{\mu\nu\rho\sigma}\partial_{\nu}F_{\rho\sigma}(x)$,
which is unrelated to the Chern classes.
The current can be shown to vanish
if the bound (3.5) holds and the problem is thus avoided. 
In general, however, monopoles cannot be excluded
and it is then possible to construct a case where eq.~(1.4) fails.

Another comment which should be made at this point is
that gauge covariant lattice Dirac operators satisfying
the Ginsparg-Wilson relation are not easy to find.
An explicit and relatively simple solution has been proposed by 
Neuberger [\ref{NeubergerI}], but so far 
the locality and differentiability of the operator has
only been proved for gauge fields with small field tensor [\ref{Locality}] 
(a sufficient condition is that eq.~(3.5) holds with $\epsilon=\frac{1}{30}$).
In particular, if one is interested in a 
rigorous construction of chiral gauge theories
using Neuberger's operator, a local gauge action should be chosen
which restricts the functional integral to this set of fields.
From a purely theoretical point of view such an action is 
perfectly acceptable, because the locality and the symmetries of the 
theory are preserved.

\section 4. Local composite fields

Since the bound (3.5) does not allow each link variable
to be varied completely independently of the other variables,
it may not be totally obvious what exactly is meant by
a local composite field. Moreover a useful definition of this term
should take into account that the locality of Neuberger's operator
(and thus of the associated topological density) is only
guaranteed up to exponentially decaying tails [\ref{Locality}].

A composite field $\phi(x)$ which depends
on the gauge field variables in a neighbourhood
of $x$ only is referred to as {\it strictly local}.
More precisely, we require that $\phi(x)$ is a smooth function of the 
link variables in a finite rectangular block 
of lattice points centred at $x$. This function should be
defined for all gauge fields on the block with
field tensor bounded by $\epsilon$. 
In the following local fields will always be assumed
to transform as a scalar field under lattice translations.

The local fields that one encounters in the context of 
Neuberger's solution of the Ginsparg-Wilson relation
may be written as a series 
$$
  \phi(x)=\sum_{k=1}^{\infty}\,\phi_k(x)
  \eqno\enum
$$
of strictly local fields $\phi_k(x)$ which are localized on blocks
$\block_k(x)$ with side-lengths proportional to $k$\/
[\ref{Locality}].
Moreover these fields and the derivatives 
$\phi_k(x;y_1,\nu_1;\ldots;y_m,\nu_m)$
with respect to the gauge field variables 
$U(y_1,\nu_1),\ldots,U(y_m,\nu_m)$
are bounded by
$$
  |\phi_k(x;y_1,\nu_1;\ldots;y_m,\nu_m)|
  \leq C_m k^{p_m}\rme^{-\theta k}, 
  \eqno\enum
$$
where the constants $C_m$, $p_m\geq0$ and $\theta>0$ can be chosen to be 
independent of the gauge field.
Such fields are local up to exponentially decaying tails.
From the point of view of the continuum limit they are as good as 
strictly local fields, since the effective localization range 
is a fixed multiple of the lattice spacing and thus microscopically
small compared to the physical distances [\ref{Locality},\ref{Niedermayer}].

\section 5. Gauge fields and vector potentials

On the lattice the link variables are the fundamental 
degrees of freedom and a vector potential is usually only introduced
in perturbation theory to parametrize the gauge fields in a 
neighbourhood of the classical vacuum configuration.
In general when one attempts to associate a vector potential 
to a given link field one runs into para\-me\-trization ambiguities.
Abelian theories are exceptional in this respect
since it is possible to pass to an equivalent description of the
theory entirely in terms of vector fields.
This has certain technical advantages which will be exploited
later.

In the context of chiral gauge theories 
the problem of how to associate a
vector potential to a given link field 
has previously been addressed in refs.~[\ref{IntA}--\ref{IntE}]. 
The following lemma extends 
a construction initially described by Hern\'andez and Sundrum
in two dimensions [\ref{IntE}].

\proclaim Lemma 5.1.
Suppose $U(x,\mu)$ is an admissible field as described sect.~3.
Then there exists a vector field $A_{\mu}(x)$ such that
$$
  \eqalignno{
  &U(x,\mu)=\rme^{iA_{\mu}(x)},
  \qquad
  |A_{\mu}(x)|\leq\pi\left(1+8\|x\|\right),
  &\enum\cr
  \noalign{\vskip2ex}
  &F_{\mu\nu}(x)=\drv{\mu} A_{\nu}(x)-\drv{\nu} A_{\mu}(x).
  &\enum\cr}
$$
Moreover, if $\tilde{A}_{\mu}(x)$ is any other field with these 
properties we have 
$$
  \tilde{A}_{\mu}(x)=A_{\mu}(x)+\drv{\mu}\omega(x),
  \eqno\enum
$$
where the gauge function $\omega(x)$ takes values that are integer
multiples of $2\pi$.

\proof
We first show that the monopole current 
$\epsilon_{\mu\nu\rho\sigma}\partial_{\nu}F_{\rho\sigma}(x)$ vanishes.
To this end we introduce the vector potential
$$
  a_{\mu}(x)={1\over i}\ln U(x,\mu),\qquad
  -\pi<a_{\mu}(x)\leq \pi,
  \eqno\enum
$$
and then note that 
$$
  F_{\mu\nu}(x)=\drv{\mu} a_{\nu}(x)-\drv{\nu} a_{\mu}(x)+2\pi n_{\mu\nu}(x),
  \eqno\enum
$$
where $n_{\mu\nu}(x)$ is an anti-symmetric tensor field with integer values. 
Eq.~(5.5) may be rewritten in terms of differential forms and after 
applying the exterior difference operator one gets
$\rmd F=2\pi\kern0.5pt\rmd n$.
The bound (3.5) implies that 
the coefficients of the form on the left-hand side of this equation
are strictly smaller than $2\pi$. Since $n_{\mu\nu}(x)$ is integer valued,
this is only possible if $\rmd F=\rmd n=0$,
i.e.~if the monopole current vanishes.

We now construct an integer vector field $m_{\mu}(x)$ such that $\rmd m=n$.
Without loss we may impose a complete axial gauge where
$m_1(x)=0$, $m_2(x)|_{x_1=0}=0$, and so on.
The non-zero components of the field are then obtained by solving
$$
  \drv{\mu}m_{\nu}(x)=n_{\mu\nu}(x)
  \quad\hbox{at}\quad x_1=\ldots=x_{\mu-1}=0
  \eqno\enum
$$
for $\mu=3,2,1$ (in this order) and $\nu>\mu$.
Using these equations and $\rmd n=0$, 
it is straightforward to verify that $\rmd m=n$
and the bound $|m_{\mu}(x)|\leq4\|x\|$ can be established
with only little more work.
The total field 
$$
  A_{\mu}(x)=a_{\mu}(x)+2\pi m_{\mu}(x)
  \eqno\enum
$$
thus has all the required properties.
The second statement made in the lemma 
may be proved by noting that
$\tilde{A}_{\mu}-A_{\mu}$ is equal to $2\pi$ times an integer vector field
with vanishing field tensor.
\endproof

\vskip1ex
Lemma 5.1 establishes a one-to-one correspondence between the
gauge equivalence classes of admissible fields $U(x,\mu)$ and the
gauge equivalence classes of vector fields $A_{\mu}(x)$ 
with field tensor $F=\rmd A$ bounded by $\epsilon$.
A gauge invariant field composed from the link variables 
may hence be regarded as a gauge invariant field 
depending on the vector potential and vice versa.

An important point to note here is that the locality properties 
of such fields are the same independently of whether they 
are considered to be functions of the link variables or the 
vector potential. Since the mapping
$$
  A_{\mu}(x)\to U(x,\mu)=\rme^{iA_{\mu}(x)}
  \eqno\enum
$$
is manifestly local, this is immediately clear if one starts with
a field composed from the link variables.
In the other direction,
starting from a gauge invariant local field $\phi(y)$
depending on the vector potential,
the key observation is that 
one is free to change the gauge of the integer 
field $m_{\mu}(x)$ in eq.~(5.7). 
In particular, we may impose a complete axial gauge taking the 
point $y$ as the origin. Around $y$
the vector potential is then locally constructed from the given link field
and $\phi(y)$ thus maps to a local function of the link variables residing
there.

The bottom-line is that, as far as gauge invariant local fields are 
concerned, we may just as well consider them to be functions of the
vector potential. 
This point of view will be adopted throughout the next section.
The gauge invariant fields that 
are being constructed may then always be rewritten 
as local functions of the link variables if so desired.

\section 6. Proof of eq.~(1.4)

We now consider a gauge invariant local field $q(x)$ which is 
defined for all admissible gauge field configurations and which is 
topological in the sense that
$$
  \sum_x\,{\partial q(x)\over\partial A_{\nu}(y)}=0.
  \eqno\enum
$$
Our aim is then to establish eq.~(1.4), 
where $\alpha$, $\beta_{\mu\nu}$ and $\gamma$ are constants and
$k_{\mu}(x)$ a gauge invariant local current.
Local here means strictly local or local with exponentially
decaying tails. The theorem holds in both cases, with $k_{\mu}(x)$
having the same locality properties as the topological field.
For simplicity the proof of eq.~(1.4) 
will be given for strictly local fields, but
the argumentation carries over almost literally 
if $q(x)$ has exponentially decaying tails.

\proclaim Lemma 6.1.
There exist gauge invariant local fields $\phi_{\mu\nu}(x)$
and $h_{\mu}(x)$ such that 
$$
  \eqalignno{
  \phi_{\mu\nu}&=-\phi_{\nu\mu},
  \qquad
  \drvstar{\mu}\phi_{\mu\nu}(x)=0,
  &\enum\cr
  \noalign{\vskip2ex}
  q(x)&=\A+\phi_{\mu\nu}(x)F_{\mu\nu}(x)+\drvstar{\mu}h_{\mu}(x).
  &\enum\cr}
$$

\proof
The vector potential $A_{\mu}(x)$ represents an admissible field
and the associated field tensor is hence bounded by $\epsilon$.
This property is preserved if 
the potential is scaled by a factor $t$ in the range $0\leq t\leq1$,
i.e.~we can contract the vector potential to zero without leaving
the space of admissible fields.
Differentiation and integration with respect to $t$ then yields
$$
  q(x)=\A+
  \sum_y\,j_{\nu}(x,y)A_{\nu}(y),
  \eqno\enum
$$
where $\A$ is equal to the value of $q(x)$ at vanishing potential and 
$$
  j_{\nu}(x,y)=\int_0^1\rmd t\,
  \left({\partial q(x)\over\partial A_{\nu}(y)}\right)_{A\to tA}.
  \eqno\enum
$$
Evidently $j_{\nu}(x,y)$ vanishes if $y$ is not contained in the 
localization region of $q(x)$ and the sum in eq.~(6.4) is hence finite.
Moreover, as a function of the gauge field, the current has the 
same locality properties as the topological field.

Since the gauge group is abelian, the derivative of 
a gauge invariant field with respect to the vector potential 
is gauge invariant and the same is hence true for $j_{\nu}(x,y)$.
Performing an infinitesimal gauge transformation in eq.~(6.4), it 
then follows that
$$
  j_{\nu}(x,y)\ldrvstar{\nu}=0.
  \eqno\enum
$$
Here and below the convention is adopted that 
a difference operator refers to $x$ or $y$ depending on whether it 
appears on the left or the right of an expression.

The Poincar\'e lemma now allows us to 
conclude that there exists a gauge invariant anti-symmetric tensor field 
$\theta_{\nu\rho}(x,y)$ such that 
$$
  j_{\nu}(x,y)=\theta_{\nu\rho}(x,y)\ldrvstar{\rho}.
  \eqno\enum
$$
As explained in sect.~2, the construction of this
field involves a reference point which is here taken to be $x$.
This choice ensures that $\theta_{\nu\rho}(x,y)$
transforms as a scalar field under lattice translations and
that it has the same locality properties as $j_{\nu}(x,y)$.

When eq.~(6.7) is inserted in eq.~(6.4), a partial summation yields
$$
  q(x)=\A+\frac{1}{2}\sum_y\,\theta_{\mu\nu}(x,y)F_{\mu\nu}(y).
  \eqno\enum
$$
This may be rewritten in the form
$$
  q(x)=\A+\phi_{\mu\nu}(x)F_{\mu\nu}(x)+
  \frac{1}{2}\sum_y\,\eta_{\mu\nu}(x,y)F_{\mu\nu}(y),
  \eqno\enum
$$
where the new fields are given by
$$
  \eqalignno{
  \phi_{\mu\nu}(x)&=\frac{1}{2}\sum_z\,\theta_{\mu\nu}(z,x),
  &\enum\cr
  \noalign{\vskip2ex}
  \eta_{\mu\nu}(x,y)&=
  \theta_{\mu\nu}(x,y)-\delta_{x,y}\sum_z\,\theta_{\mu\nu}(z,y).
  &\enum\cr}
$$
Both of them are gauge invariant and anti-symmetric in 
the indices $\mu,\nu$. 
Moreover, taking the locality properties of $\theta_{\mu\nu}(x,y)$
into account, it is easy to prove that $\phi_{\mu\nu}(x)$ is a local
field.

To complete the proof of the lemma we now need to show
that $\drvstar{\mu}\phi_{\mu\nu}(x)=0$ and 
that the last term in eq.~(6.9) is equal to $\drvstar{\mu}h_{\mu}(x)$,
where $h_{\mu}(x)$ is a gauge invariant local current.
Using eq.~(6.7) and the anti-symmetry of the tensor field
one obtains
$$
  \drvstar{\mu}\phi_{\mu\nu}(x)=-\frac{1}{2}\sum_z\,j_{\nu}(z,x).
  \eqno\enum
$$
From eq.~(6.1) and the definition (6.5)
it is then immediately clear that the 
right-hand side of this equation vanishes.
The field $\phi_{\mu\nu}(x)$ thus has all the properties
stated in lemma. 

As for the other field we note that
$$
  \sum_x\,\eta_{\mu\nu}(x,y)=0
  \eqno\enum
$$
and the Poincar\'e lemma may hence be applied again, this time 
with reference point~$y$.
This leads to the representation
$$
  \eta_{\mu\nu}(x,y)=\drvstar{\lambda}\tau_{\lambda\mu\nu}(x,y)
  \eqno\enum
$$
in terms of a new field $\tau_{\lambda\mu\nu}(x,y)$.
In particular, the divergence of the local current
$$
  h_{\mu}(x)=\frac{1}{2}\sum_y\,\tau_{\mu\nu\rho}(x,y)F_{\nu\rho}(y)
  \eqno\enum
$$
is equal to the last term in eq.~(6.9) and the lemma has thus been proved.
\endproof

\vskip1ex
In the second step of the proof of eq.~(1.4) we determine
the general form of the field $\phi_{\mu\nu}(x)$ 
using no other properties than those stated in lemma~6.1.
Compared to our argumentation above, there is very
little difference here. The algebra is more involved, however,
and will be presented in full detail.

\proclaim Lemma 6.2.
There exists a gauge invariant, local and totally anti-symmetric 
tensor field $t_{\lambda\mu\nu}(x)$ such that
$$
  \phi_{\mu\nu}(x)=\B_{\mu\nu}+
  \C\epsilon_{\mu\nu\rho\sigma}F_{\rho\sigma}(x+\hat{\mu}+\hat{\nu})+
  \drvstar{\lambda}t_{\lambda\mu\nu}(x).
  \eqno\enum
$$

\proof
Proceeding as in the proof of lemma~6.1, 
it is straightforward to derive 
a representation analogous to eq.~(6.8)
for the field $\phi_{\mu\nu}(x)$. 
Only the locality and gauge invariance of the field
are required for this and one ends up with the expression
$$
  \phi_{\mu\nu}(x)=\beta_{\mu\nu}
  +\frac{1}{2}\sum_y\,\xi_{\mu\nu\rho\sigma}(x,y)F_{\rho\sigma}(y),
  \eqno\enum
$$
where $\beta_{\mu\nu}$ is equal to the value 
of $\phi_{\mu\nu}(x)$ at vanishing gauge potential.
The new field appearing in eq.~(6.17) satisfies
$$
  \xi_{\mu\nu\rho\sigma}=-\xi_{\nu\mu\rho\sigma}=-\xi_{\mu\nu\sigma\rho},
  \qquad
  \drvstar{\mu}\,\xi_{\mu\nu\rho\sigma}(x,y)\ldrvstar{\rho}=0.
  \eqno\enum
$$
It is gauge invariant and has the same locality properties 
as the current $j_{\nu}(x,y)$ that we have discussed earlier.
The fields introduced in the following lines are also of this 
type, but for brevity we shall not mention this again.

Starting from eq.~(6.18) and applying the Poincar\'e lemma two times 
we have
$$
  \eqalignno{
  &\drvstar{\mu}\,\xi_{\mu\nu\rho\sigma}(x,y)=
  \upsilon_{\nu\lambda}(x,y)\ldrvstar{\tau}\,\epsilon_{\lambda\tau\rho\sigma},
  &\enum\cr
  \noalign{\vskip2ex}
  &\drvstar{\nu}\,\upsilon_{\nu\lambda}(x,y)=
  \omega(x,y)\ldrvstar{\lambda},
  &\enum\cr}
$$
where $x$ has been taken as the reference point
when constructing the fields $\upsilon_{\nu\lambda}(x,y)$ and $\omega(x,y)$.
An immediate consequence of the last equation is that 
$$
  \C=-\frac{1}{2}\sum_z\,\omega(z,x)
  \eqno\enum
$$
is independent of $x$. Moreover, in view of the locality properties 
of the expression, a dependence on the gauge field is then also excluded 
and $\gamma$ is hence a constant.

Another application of the Poincar\'e lemma
now implies that 
$$
  \omega(x,y)=-2\C\delta_{x,y}+\drvstar{\nu}\varphi_{\nu}(x,y)
  \eqno\enum
$$
for some vector field $\varphi_{\nu}(x,y)$. If we define
$$
  \hat{\upsilon}_{\nu\lambda}(x,y)=
  \upsilon_{\nu\lambda}(x,y)-\varphi_{\nu}(x,y)\ldrvstar{\lambda},
  \eqno\enum
$$
it is then straightforward to prove the relations
$$
  \eqalignno{
  &\drvstar{\mu}\,\xi_{\mu\nu\rho\sigma}(x,y)=
  \hat{\upsilon}_{\nu\lambda}(x,y)
  \ldrvstar{\tau}\,\epsilon_{\lambda\tau\rho\sigma},
  &\enum\cr
  \noalign{\vskip2ex}
  &\drvstar{\nu}\,\hat{\upsilon}_{\nu\lambda}(x,y)=
  -2\C\delta_{x,y}\ldrvstar{\lambda}.
  &\enum\cr}
$$
Compared to eqs.~(6.19),(6.20) the important difference is that 
the form of the right-hand side of the second equation is 
now known precisely.

In the next step we propagate this information to the first equation
by noting that 
$$
  \delta_{x,y}\ldrvstar{\lambda}=
  -\drvstar{\nu}\left\{\delta_{\nu\lambda}\delta_{x,y-\hat{\nu}}\right\}.
  \eqno\enum
$$
The general solution of eq.~(6.25) is hence given by
$$
  \hat{\upsilon}_{\nu\lambda}(x,y)=2\C\delta_{\nu\lambda}\delta_{x,y-\hat{\nu}}
  +\drvstar{\mu}\vartheta_{\mu\nu\lambda}(x,y),
  \qquad
  \vartheta_{\mu\nu\lambda}=-\vartheta_{\nu\mu\lambda}.
  \eqno\enum
$$
It follows from this that the shifted field
$$
  \hat{\xi}_{\mu\nu\rho\sigma}(x,y)=
  \xi_{\mu\nu\rho\sigma}(x,y)-
  \vartheta_{\mu\nu\lambda}(x,y)\ldrvstar{\tau}\,
  \epsilon_{\lambda\tau\rho\sigma}
  \eqno\enum
$$
satisfies the relations
$$
  \eqalignno{
  &\phi_{\mu\nu}(x)=\beta_{\mu\nu}
  +\frac{1}{2}\sum_y\,\hat{\xi}_{\mu\nu\rho\sigma}(x,y)F_{\rho\sigma}(y),
  &\enum\cr
  \noalign{\vskip1.5ex}
  &\drvstar{\mu}\,\hat{\xi}_{\mu\nu\rho\sigma}(x,y)=
  2\C\delta_{x,y-\hat{\nu}}\ldrvstar{\tau}\,
  \epsilon_{\nu\tau\rho\sigma}.
  &\enum\cr}
$$
We may now again use the identity~(6.26) 
and the Poincar\'e lemma to infer that 
$$
  \hat{\xi}_{\mu\nu\rho\sigma}(x,y)=
  2\C\epsilon_{\mu\nu\rho\sigma}\delta_{x,y-\hat{\mu}-\hat{\nu}}
  +\epsilon_{\mu\nu\lambda\tau}\drvstar{\lambda}
  \kappa_{\tau\rho\sigma}(x,y),
  \eqno\enum
$$
where $\kappa_{\tau\rho\sigma}(x,y)$ is another tensor field.
Together with eq.~(6.29) and the definition 
$$
  t_{\lambda\mu\nu}(x)=\frac{1}{2}\sum_y\,\epsilon_{\lambda\mu\nu\tau}
  \kappa_{\tau\rho\sigma}(x,y)
  F_{\rho\sigma}(y),
  \eqno\enum
$$
this proves the lemma. \endproof

\vskip1ex
The combination of lemma 6.1 and lemma 6.2 leads to the representation
$$
  \eqalignno{
  q(x)&=\A+\B_{\mu\nu}F_{\mu\nu}(x)+
  \C\epsilon_{\mu\nu\rho\sigma}
  F_{\mu\nu}(x)F_{\rho\sigma}(x+\hat{\mu}+\hat{\nu})+\drvstar{\mu}h_{\mu}(x)
  &\cr
  \noalign{\vskip2ex}
  &\quad
  +\drvstar{\mu}t_{\mu\nu\rho}(x)F_{\nu\rho}(x).
  &\enum\cr}
$$
Only the last term in this equation
does not seem to fit with eq.~(1.4), but using 
the anti-symmetry of the tensor field $t_{\lambda\mu\nu}(x)$ 
and the vanishing of the monopole current, 
$\epsilon_{\mu\nu\rho\sigma}\partial_{\nu}F_{\rho\sigma}(x)=0$,
it is easy to check that
$$
  \drvstar{\mu}t_{\mu\nu\rho}(x)F_{\nu\rho}(x)=
  \drvstar{\mu}\bigl\{t_{\mu\nu\rho}(x)F_{\nu\rho}(x+\hat{\mu})\bigr\},
  \eqno\enum
$$
i.e.~this is a contribution to the divergence term $\drvstar{\mu}k_{\mu}(x)$
and the proof of eq.~(1.4) is thus complete.

In principle the construction presented in this section provides
an explicit expression for the current $k_{\mu}(x)$ in the form
of a certain linear combination of the first and second derivative
of the topological field with respect to the gauge potential.
No attempt has here been made to work this out, because the resulting
expression tends to be very complicated and is thus unlikely to 
be of any practical use.
Maybe a more elegant formula can be found 
now that eq.~(1.4) is known to be true.

\section 7. Concluding remarks

For the case of abelian lattice gauge theories, 
and if only gauge fields satisfying the bound (3.5) are considered,
the theorem proved in this paper shows that 
there are no topological fields other than those associated 
with the Chern classes. 
This complements the geometrical constructions of the 
topological charge in these theories
[\ref{TopA},\ref{TopB}] and makes it evident that the
charge assignment is unique for admissible fields. 
To establish a similar theorem 
in non-abelian lattice gauge theories
however seems to be considerably more difficult. 

The constant $\gamma$ in eq.~(1.4)
can be worked out straightforwardly by expanding the topological
field in powers of the vector potential.
In the case of the axial anomaly (1.2) the 
constant has to be of the form
$$
  \gamma={n\over32\pi^2},\qquad n\in\gz,
  \eqno\enum
$$
as otherwise one would run into a contraction with the 
exact index theorem of ref.~[\ref{HasenfratzEtAl}].
If the lattice Dirac operator $D$ describes a single fermion with 
charge $\rme=1$ and if there are no doubler modes in 
the free fermion limit, one expects that 
$n=\pm1$ (depending on conventions).
This has been confirmed by explicit calculation for 
the perfect Dirac operator
[\ref{GinspargWilson}--\ref{HasenfratzEtAl}]
and now also for Neuberger's operator [\ref{KikukawaYamada}]. 
However, as already emphasized by Chiu [\ref{Chiu}], there is
currently no general theorem which guarantees that $n$ has the 
correct value for any decent choice of $D$.

Eq.~(1.4) holds independently of the transformation behaviour
of the topological field under lattice rotations.
If $q(x)$ transforms as a pseudo-scalar,
it is easy to show that the first two terms on the 
right-hand side of the equation have to vanish.
The remaining terms individually do not transform as a pseudo-scalar field,
but one can enforce this by
averaging the equation over the group of lattice rotations
with the appropriate weight factors so as to project on the pseudo-scalar
component.

As recently reported by Niedermayer [\ref{Niedermayer}],
left- and right-handed lattice fermion fields may
be introduced in a natural way if the lattice Dirac operator satisfies 
the Ginsparg-Wilson relation.
The gauge anomaly then shows up when one attempts
to define the associated functional integration measure.
In abelian theories with anomaly-free multiplets of Weyl fermions
it now appears to be possible to construct a measure 
which preserves the gauge symmetry and the locality of the theory.
One of the problems which one has here is that
the topology of the field space 
can give rise to global obstructions which are not necessarily connected
with the local anomaly (the $\gz_2$ anomaly in 
$\SUtwo$ gauge theories [\ref{Witten},\ref{NeubergerII}] 
is an example for this).

\vskip1ex
I would like to thank 
Peter Hasenfratz, Ferenc Niedermayer and Peter Weisz
for correspondence and many interesting discussions
on the problem of how to put chiral gauge theories on the lattice.
I am also indebted to Pilar Hern\'andez and Ting-Wai Chiu for 
their comments on the first version of this paper.

\appendix A

All fields considered in this paper live on a 
hyper-cubic euclidean lattice
with lattice spacing $a=1$ and infinite extent in all directions.
Except in section~2
the dimension $n$ of the lattice is equal to four.
Lorentz indices $\mu,\nu,\ldots$ run from $1$ to $n$.
Throughout the paper 
the Einstein summation convention is applied to these indices.
The symbol $\epsilon_{\mu\nu\rho\sigma}$ stands for the totally anti-symmetric
tensor in four dimensions with $\epsilon_{1234}=1$.
In any dimension $\delta_{x,y}$ is equal
to $1$ if $x=y$ and zero otherwise. 

The forward and backward nearest-neighbour difference operators 
act on lattice functions $f(x)$ according to 
$$
  \eqalignno{
  \drv{\mu}f(x)&=f(x+\hat{\mu})-f(x),
  &\enum\cr
  \noalign{\vskip2ex}
  \drvstar{\mu}f(x)&=f(x)-f(x-\hat{\mu}),
  &\enum\cr}
$$
where $\hat{\mu}$ denotes the unit vector in direction $\mu$.
The operators $\ldrv{\mu}$ and $\ldrvstar{\mu}$ 
which may appear on the right of a function $f(x,y)$ are defined in 
exactly the same way but refer to the second argument $y$.


\beginbibliography

\bibitem{GinspargWilson}
P. H. Ginsparg and K. G. Wilson,
Phys. Rev. D25 (1982) 2649

\bibitem{HasenfratzI}
P. Hasenfratz,
Nucl. Phys. B (Proc. Suppl.) 63A-C (1998) 53

\bibitem{HasenfratzEtAl}
P. Hasenfratz, V. Laliena and F. Niedermayer,
Phys. Lett. B427 (1998) 125

\bibitem{NeubergerI}
H. Neuberger,
Phys. Lett. B417 (1998) 141;
{\it ibid}\/ B427 (1998) 353

\bibitem{HasenfratzII}
P. Hasenfratz,
Lattice QCD without tuning, mixing and current renormalization,
hep-lat/9802007

\bibitem{LuscherI}
M. L\"uscher,
Phys. Lett. B428 (1998) 342

\bibitem{KikukawaYamada}
Y. Kikukawa and A. Yamada,
Weak coupling expansion of massless QCD with 
a Ginsparg-Wilson fermion and axial U(1) anomaly,
hep-lat/9806013

\bibitem{BottTu}
R. Bott and L. W. Tu, Differential forms in algebraic topology
(Springer-Verlag, New York, 1982)

\bibitem{Bertlmann}
R. A. Bertlmann, Anomalies in quantum field theory (Oxford University
Press, Oxford, 1996)

\bibitem{Locality}
P. Hern\'andez, K. Jansen and M. L\"uscher,
Locality properties of Neuberger's lattice Dirac operator,
hep-lat/9808010

\bibitem{Niedermayer}
F. Niedermayer,
Exact chiral symmetry, topological charge and related topics,
plenary talk given at the International Symposium on Lattice Field Theory,
Boulder, July 13-18, 1998

\bibitem{IntA}
M. G\"ockeler, A. Kronfeld, G. Schierholz and U. J. Wiese,
Nucl. Phys. B404 (1993)

\bibitem{IntB}
P. Hern\'andez and R. Sundrum,
Nucl. Phys. B455 (1995) 287

\bibitem{IntC}
G. 't Hooft,
Phys. Lett. B349 (1995) 491

\bibitem{IntD}
G. T. Bodwin,
Phys. Rev. D54 (1996) 6497

\bibitem{IntE}
P. Hern\'andez and R. Sundrum,
Nucl. Phys. B472 (1996) 334

\bibitem{TopA}
M. L\"uscher,
Commun. Math. Phys. 85 (1982) 39

\bibitem{TopB}
A. V. Phillips and D. A. Stone,
Commun. Math. Phys. 103 (1986) 599;
{\it ibid}\/ 131 (1990) 255

\bibitem{Witten}
E. Witten,
Phys. Lett. B117 (1982) 324

\bibitem{NeubergerII}
H. Neuberger,
Witten's SU(2) anomaly on the lattice,
hep-lat/9805027

\bibitem{Chiu}
T. W. Chiu,
The axial anomaly of Ginsparg-Wilson fermion,
hep-lat/9809013

\endbibliography

\bye